\documentclass[12pt]{article}
\usepackage[utf8]{inputenc}
\usepackage{hyperref}
\usepackage{amsmath}
\usepackage{amssymb}
\usepackage{amsfonts}
\usepackage{bbm}
\usepackage{cite}
\usepackage{pifont}
\usepackage{mathtools}
\usepackage{esint}

\title{Synthetic Dynamics}
\author{VS Morales-Salgado}
\date{}
\begin{document}

\maketitle

\begin{abstract}
    This work reflects on mechanics as an epistemological framework on the state of a physical system to regard dynamics as the distribution of mechanical properties over spacetime coordinates.
    The resulting distribution is taken to be the partition function of the relevant physical quantities over a spacetime parametrized by coordinates.
    The partition yields a probabilistic interpretation that, based on Feynman's path integral formulation, leads to a dynamical law that generalizes the Schr\"odinger equation.
    A variety of systems can be put into the form proposed here, including particles in potentials, as well as matter and interaction fields.
    The main advantage of the proposed framework is that it presents the space of properties separately from that of the space of coordinates, whereas the dynamical law can be interpreted as the equation of two differential structures, one from each of these spaces.
    The resulting framework shows possibilities to further study physical quantities that relate directly to the spacetime coordinates, whose dynamics is best described in thermodynamical, rather than Hamiltonian, terms.
    A notable example is the theory of general relativity, in which the case of a scalar field in a Robertson-Walker metric is explored.
   
\end{abstract}

\section{Introduction}
    Discovering the nature of a system implies making assertions about its properties to be corroborated against observations.
    Thus, we regard modern mechanical theories as assertions about the mechanical quantities that describe a system.
    Note that this is an epistemological, rather than ontological, approach \cite{b18}; which helps embedding uncertainty into it more naturally and readily yields a probabilistic interpretation of mechanics.
    For that reason, dealing with distributions of physical quantities, rather than those quantities directly, can lead to advantages in formulating statements about the system of interest.
    At the same time, there must be a limiting case, known as a macroscopic limit, that tends towards a minimization of uncertainty \cite{u76}.
   
    In this set-up, dynamics can be regarded as a statement about how that distribution propagates with respect to spacetime coordinates.
    Thus, dynamical statements necessarily distinguish two sets of quantities: system properties and spacetime coordinates; the simplest being an equation between changes with respect to properties and changes with respect to spacetime.
    This results in a straightforward interpretation of properties as fields over spacetime.
   
    This investigation builds on several ideas that have been put forward in physics.
    We note that two physical theories that directly deal with uncertainty and thus rely heavily on probabilistic interpretations are statistical mechanics \cite{k16} and quantum mechanics \cite{m49}.
    Thus, here we shall take from both to build a dynamical framework that remains useful for physical investigations.
    Specifically, we identify the distribution of mechanical properties with the partition function in statistical mechanics and the path integral in quantum mechanics \cite{w13}, as well as their formal comparison through the procedure known as Wick rotation \cite{s15}.
   
    From quantum mechanics, we also exploit the correspondence principle towards classical mechanics \cite{v28}, noting that it is particularly tractable when passing from the Schr\"odinger equation to the Hamilton-Jacobi equation \cite{j25}.
    A procedure also known as the WKB approximation \cite{h13}.
    This is key because in both of those equations the role of the two sets of quantities (system properties and spacetime coordinates) are clearly distinguished with respect to a distribution function: the propagator and the principal function, respectively.
    One could say that the synthetic dynamics proposed here explores the covariant dynamics of systems that can be approximated using the WKB method.
    Thus, in the macroscopic limit the resulting dynamical law is akin to the so-called De Donder-Weyl formulation of the Hamilton-Jacobi equation \cite{h35}.
   
    On the other hand, statistical mechanics presents a more transparent epistemological interpretation that derives from a more practical approach towards uncertainty \cite{s86}.
    From this theory, we exploit its correspondence with thermodynamics as an emergent phenomena, that is, with no straightforward microscopic analog \cite{aks74}.
    This is crucial for theories like general relativity that face trivialities when dealt using the Hamiltonian formalism \cite{uw89}.
   
    It is also worth to emphasize that, although this is an epistemological approach to mechanical theories, this work is mostly dedicated to the narrower problem of establishing dynamical laws consistent with that approach.
    Thus, for the specific case of quantum mechanics, no modification of its postulates is presented.
    If at all, this work deals with the time evolution postulate by generalizing the Schr\"odinger equation as put forward \cite{s26}.
    Something similar can be said about statistical mechanics, although it can be developed using either classical or quantum mechanics as a fundamental theory.
    Here we are rather concerned with the statement and solution of their dynamical problem in terms of how their relevant quantities distribute in a coordinate space.
   
    Consequently, the objective of this work is to generalize mechanical concepts as epistemological devices to know about a system so that the framework fits those observations about the dynamical behavior of the quantities involved.
    Along with the presentation of the framework, some of its features are described and its usefulness to study certain systems of general interest is tested.
   
    While the proposed view of fundamental mechanics is indeed general, the requirement that it contains a correspondence and emergence with a so-called macroscopic behavior demands particular conditions on the dynamical laws.
    Then, rather than focusing on the quantization, one is interested in the opposite procedure that in more algebraic-prone studies is usually dubbed a deformation of the corresponding structure.
    This is the reason behind the denomination \emph{synthetic}: the epistemological approach allows one to focus on effectively reproducing the observed dynamical laws disregarding the ontological nature of the physical quantities.
    This is only possible by means of the shift resulting from the historical development of statistical and quantum mechanics that forced to incorporate uncertainty into physical thought.
   
    The article is organized as follows:
    In section \ref{GME}, the epistemological interpretation of mechanics is presented.
    It takes the partition as the central object and yields the first equation analogous to the fundamental thermodynamic relation.
    A general dynamical law is obtained by demanding it holds a specific correspondence principle.
    In section \ref{SS}, some consequences of the dynamical law are established by applying it to different classes of systems.
    These systems include particle systems, where the properties are mainly positions over a single coordinate commonly identified with time, as well as fields, where several coordinates are taken into account.
    In section \ref{CA}, a final case is presented and investigated.
    It shows the possibility to use a complex action to have a system that corresponds to classical fields plus a property that relates directly to the coordinates through an entropy function and whose macroscopic behavior is obtained non-trivially by thermodynamical reasoning, rather than Hamiltonian one.
    Thus, it leads to a proposal for how general relativity comes to emerge from more fundamental mechanics.
    As an example, we explore the case of a scalar field in a Robertson-Walker metric.
    Finally, in section \ref{CC}, the concluding remarks are given and some prospective work is stated.
           
\section{Generalizing mechanical epistemology}\label{GME}
    The idea behind a \emph{synthetic dynamics} is to generalize the way one uses statistics and probability to understand the dynamical laws of properties of systems in terms of how they depend on spacetime coordinates.
    Thus, a given physical system possesses mechanical properties, denoted here by $\xi=\left\{\xi_i\right\}$, $i=1,...,n$.
    To know about the state of the system, one must collect \emph{observations} of those properties,  that, in turn, result in associated values $\xi_c$ that one supposes can be described in the following terms:
        \begin{equation}\label{MeaCon}
         \mathcal{M}(\xi_c) = 0 \,.
        \end{equation}
    This is written in the form of a \emph{constraint} since that is effectively the epistemic function of observations: up to uncertainties, whether fundamental or practical, the measured values are the information one has confirmed about the system through corroboration by the act of measuring.  
    Complementarily, what happens to the system in between observations is unknown.
   
    Then, a particular system is characterized by the relations among its properties $\xi$.
    These are inferred from the results of the observations $\xi_c$ and they yield eq. (\ref{MeaCon}).
    However, in general, when it is not observed, the system (or rather, its properties) need not satisfy eq. (\ref{MeaCon}).
    In that case, the relations among properties $\xi$ can only be probabilistic.
   
    Now, one expects to have a quantity that incorporates all the possible relations among the components of $\xi$, assigning a weight value to each of those \emph{configurations}.
    While one can imagine several options for that quantity, here we consider the case that the particular epistemological description of the system is then specified by its \emph{action}:
        \begin{equation}
         \mathcal{S} = \mathcal{S}\left(\xi_c\right) = \mathcal{S}\left[\xi\right] \,,
        \end{equation}
    where we try to make it clear that $\mathcal{S}$ is a function of $\xi_c$, but a functional of $\xi$, by means of the use of distinct brackets.
   
    Thus far, one can make statements about how $\mathcal{S}$ is distributed over the different possible values $\xi$ can sample.
    In fact, this is where richer structures, e.g., symmetries, may be included in the set of $\xi$.
    However, distinct system behaviors imply different relations among their properties $\xi$.
    These differences are captured mathematically by parameters whose values distinguish one system from another.
   
    Dynamics then is the behavior of $\mathcal{S}$ and $\xi$ over a certain set of parameters $\chi=\left\{\chi_j\right\}$, $j=1,...,d$.
    What is particular about these parameters is that they form a space with the structure and characteristics that allow one to talk about position, orientation and distance.
    Thus described, one can make $\chi$'s structure more rigid or flexible depending on observations.
    One immediate result of such a perspective is that it embodies the notion of parametrization invariance.
    In contemporary physics language, $\chi$ can be naturally regarded as a set of coordinates of a given spacetime.
   
    Both variations of $\mathcal{S}$ with respect to $\xi$ and $\chi$ have been extensively studied.
    Here, we wish to use all those results.
    First, the local behavior of $\mathcal{S}$, with respect to $\chi$, yields the \emph{Lagrangean} $\mathcal{L}$ in the sense that:
        \begin{equation}\label{ActLag}
         \mathcal{S} = \int_\mathcal{C} \mathcal{L} \,,
        \end{equation}
    where $\mathcal{C}$ describes the boundary conditions and $\mathcal{L}$ may be interpreted as a \emph{differential action}.
    However, the limits of integration must be in agreement with the constraints defined by eq. (\ref{MeaCon}).
    That is the classical interpretation of the action that is also exploited in the path integral formulation of quantum mechanics.
   
    Thus, the observed values are incorporated as constraints through the boundary of integral (\ref{ActLag}).
    Outside of observations, all possible configurations must be accounted for.
    This is precisely what the action $\mathcal{S}$ does.
    Nevertheless, $\mathcal{S}$ must be inferred from the relations among $\xi$ obtained from observations, i.e. $\xi_c$.
 
    Now that we established that $\mathcal{S}$ encodes the mechanical information of the system, we seek to relate it to a formulation that is operational for a sequence of probabilistic observations.
    Accordingly, we assume that the weight assigned to configurations $\xi$ compatible with the observations $\xi_c$ given by:
        \begin{equation}\label{mu}
         \mu(\xi_c) = \mu[\xi] = \exp\left( -\kappa\,\mathcal{S} \right) \,\mathcal{D}\xi\,,
        \end{equation}
    where $\kappa$ is a parameter and $\mathcal{D}\xi$ is a functional differential with respect to $\xi$.
   
    This means that the propagator $K(\xi_c)$ which describes how mechanical properties compatible with observations $\xi_c$  evolve is given by a \emph{functional integral}:
        \begin{equation}
         K(\xi_c) = \fint_{\mathcal{C}} \mu \,,
        \end{equation}
    where the symbol $\fint$ denotes a functional integral with respect to the weight $\mu$ as given by (\ref{mu}).
   
    Then, the interpretation of $K$ depends on the parameter $\kappa$:
    If $\kappa$ is real and positive, $\mu$ is a real and positive measure, and $K$ propagates probability densities, like in classical statistical mechanics.
    If  $\kappa={\rm i}/\hbar$, $\mu$ is complex, and $K$ is a quantum amplitude propagator, whose squared modulus yields transition probabilities, like a Feynman propagator.
    In all cases, $K$ functions as a dynamical propagator and probabilities arise from the distributions that $K$ transports.
   
    When the set of configurations is parametrized such that $\chi_a$ and $\chi_b$ label independent degrees of freedom on $\mathcal{C}$, $K$ explicitly transports the state from $\xi_a$ at $\chi_a$ to $\xi_b$ at $\chi_b$.
    This allows us to treat $\xi$ as a field $\xi(\chi)$ and study its dynamics.
    In short, $K(\xi_c)$ is the object that makes it possible to define and compute dynamical quantities.
   
    Now, \emph{expectation values} of properties $\xi$ or, more generally of functions of them $f(\xi)$, are defined as:
        \begin{equation}
         \langle f\rangle = \frac{1}{Z} \fint f(\xi) \, \mu \,,
        \end{equation}
    where, for a shorter notation, we have fixed and omitted the relevant constraint $\mathcal{C}$, and defined the \emph{partition function} as:
        \begin{equation}\label{parti}
         Z := \fint \mu \,.
        \end{equation}

    This is an epistemological interpretation of mechanics as it allows to describe mechanical concepts in terms of what we actually know via observation, what we do not know but can describe as a possibility and compare those possibilities using probabilities.
    However, before studying specific examples, one can say some things about the immediate consequences of this framework.
   
    First, in the definition of the partition (\ref{parti}), one has (functionally) integrated out dependencies on $\xi$ and we have that $Z=Z(\xi_c)$.
    Thus, we define the \emph{free action} $F=F(\xi_c)$ as follows:
        \begin{equation}\label{FrAct}
         F := -\frac{1}{\kappa}\,\ln\,Z \,,
        \end{equation}
    where the proportionality constant is conveniently chosen as $-\frac{1}{\kappa}$, akin to the analogous treatment in statistical mechanics, so that $Z=\exp(-\kappa F)$ and one can write the normalized probability weight:
        \begin{equation}
         \varrho(\xi_c) = \varrho[\xi] = \frac{\mu}{Z}
                            = \exp\left(\kappa\,(F-\mathcal{S})\right)  \,.
        \end{equation}
    Then, we can write expectation values as $\bar{f}(\xi) = \fint f(\xi) \, \varrho\,$ and, given the linearity of functional integration, one can readily obtain the following equation:
        \begin{equation}\label{1id}
         F = \langle\mathcal{S}\rangle + \frac{\sigma}{\kappa} \,,
        \end{equation}
    where we have defined $\sigma$ similarly to the \emph{entropy} in statistical mechanics:
        \begin{equation}
         \sigma := -\fint \ln(\varrho)\,\varrho \,.
        \end{equation}
         
    \subsection{A correspondence principle}
    The previous interpretation of mechanics could do without the parameter $\kappa$.
    However, formally, its inclusion provides a scale of the uncretainty when the system is not observed; either because it is in thermal contact with a bath (epistemic uncertainty, controlled by $\kappa=1/k_B\,T$) or because quantum mechanics imposes irreducible fluctuations (ontic uncertainty, controlled by $\kappa=1/\hbar$).
   
    Along with the parameter $\kappa$, there is a limiting case ($\kappa\gg1$) where the evolution of the system occurs seamlessly on the constraining observations given by eq. (\ref{MeaCon}).
    Thus, its explicit consideration is essential for controlling the interpolation between classical, quantum, and mixed regimes, and for quantifying the relative magnitude of thermal versus quantum uncertainties in systems that are not observed at the microscopic level.
   
    For example, the case of quantum mechanics, where $\kappa = {\rm i}/\hbar$ yields the following condition as such a limiting case:
        \begin{equation}\label{PEA}
         \delta\mathcal{S} = 0 \,,
        \end{equation}
    that, in turn, leads to several equivalent formulations of classical mechanics on $\mathcal{M}(\chi_c)=0$.
    One of this is the \emph{Euler-Lagrange} equations:
        \begin{equation}
         \int_\mathcal{C}\left(\nabla\cdot\left(\partial_{\nabla\xi}\mathcal{L}\right) - \partial_\xi\mathcal{L} \right) = \int_{\partial\mathcal{C}}\partial_{\nabla\xi}\mathcal{L}\,,
        \end{equation}
    where, the boundary term in the right hand side usually vanishes.
   
    Another one being the Hamilton-Jacobi equation that prescribes classical mechanics through \emph{Hamilton's principal function} $\Omega(\xi_c,\chi)$, which can be identified with the value of $\mathcal{S}(\xi_c)$ along the path $\xi_c$ that satisfies condition (\ref{PEA}).
    Then, one can write ${\rm d}\Omega=\mathcal{L}|_{\xi_c}{\rm d}\chi$.
    On the other hand, if we compute ${\rm d}\,\Omega = (\partial_{\xi_c}\Omega\,\nabla\xi_c + \nabla\Omega){\rm d}\chi$, then we obtain:
        \begin{equation}\label{preHJ}
         \mathcal{L}|_{\chi_c} = \rho_c\nu_c + \nabla\Omega \,,
        \end{equation}
    where we have defined the \emph{velocity} $\nu_c:=\nabla\xi_c$ and the \emph{impetus} $\rho_c:=\partial_{\xi_c}\Omega$.
    The latter can be identified as the flux of generalized momentum through the boundary $\partial\mathcal{C}$, i.e. $\rho_c=\int_{\partial\mathcal{C}}\partial_{\nabla\xi}\mathcal{L}$.
    This allows us to identify the \emph{Hamiltonian} $\mathcal{H}|_{\xi_c}=\rho_c\nu_c-\mathcal{L}|_{\xi_c}$ along $\xi_c$ and one has the \emph{Hamilton-Jacobi equation}:
        \begin{equation}\label{HJe}
         \nabla\Omega|_{\partial\mathcal{C}} = -\mathcal{H}(\xi_c,\partial_{\xi_c}\Omega,\chi) \,.
        \end{equation}
    These definitions are actually generalized notions compared to the usual ones employed in classical mechanics since, instead of a single parameter identified as time, here one may have several parameters $\chi$ describing position in spacetime.
    Thus, $\mathcal{H}$ can be regarded as the generator of translations in the space of parameters $\chi$, along the direction normal to the boundary $\partial\mathcal{C}$.
    If such a direction is denoted as $n_\tau$, then one can write $\nabla\Omega|_{\partial\mathcal{C}}=n_\tau\cdot\nabla\Omega$.
   
    Since we wish that the evolution of $Z$ be connected with the dynamics of $\Omega$ in the limit $\kappa\gg1$, let us prescribe that $Z\propto\exp(\kappa\,\Omega)$.
    In fact, the constant of proportionality can still depend on $\chi$ and even on $\kappa$.
    So, we assume that the partition takes the following exponential form:
        \begin{equation}\label{canon1}
         Z = \exp\left( \Theta(\chi,\kappa) + \kappa\,\Omega(\xi_c,\chi) \right) \,.
        \end{equation}
    The parameter $\kappa$ can be used to locally approximate $Z$.
    If we recall that $Z$ can be viewed as a kernel propagating mechanical properties $\xi$ over the space of parameters $\chi$, then eq. (\ref{canon1}) can be regarded as a spacetime distribution with a dominance of $\Omega$ depending on $\kappa$; whereas the role of $\Theta$ is to counter such a dominance in the opposite case, acting as a noise term.

    The previous prescription and its corresponding results, in the light of eq. (\ref{1id}), yields the identification:
        \begin{equation}
         \Omega = \lim_{\kappa\to0} \fint \mathcal{S} \varrho \,.
        \end{equation}
    This is the \emph{correspondence principle} in terms of the action.
    Thus far, assumption (\ref{canon1}) is a specific form of the partition that allows us to connect two macroscopic descriptions: the emergence of thermodynamical quantities and the limiting case where $\kappa\to0$.
         
    \subsection{Dynamics of mechanical properties through the partition}
    In what follows we will employ the ideas previously introduced to explore the features of the partition as a propagator of the observed values $\xi_c$ over the space of parameters $\chi$.
    This is what we shall understand as the dynamics of mechanical properties $\xi$.
    Indeed, the interpretation of $\chi$ motivated here is that of a set of parameters that help define position, orientation and proximity.
    That makes them identifiable as \emph{coordinates} in a space.
    Although, thus far, we have not specified any structure of the set of parameters $\chi$ and, with it, of spacetime.
   
    In order to study the dynamics of the system, we must investigate the local behavior of $Z$, that we assume takes the form (\ref{canon1}).
    This entails finding the differential equation ruling it by proposing one to be tested against observations.
    Since $\mathcal{H}$ has been found to be the generator of translations with respect to $\chi$, it is reasonable to assume that $n_\tau\cdot\nabla_\chi Z$ and $\mathcal{H}(Z)$ are proportional, in the fashion of a Schr\"odinger equation.
    Further, we shall suppose that the direction $n_\tau$ is the one defined by the very same gradient $\nabla_\chi Z$, so that $\mathcal{H}$ naturally translates $Z$ in the direction of its greatest increase.
    Then, we shall suppose that the partition satisfies the \emph{generalized Schr\"odinger equation}:
        \begin{equation}\label{GS}
         D_\chi Z = \kappa\,\mathcal{H}(Z) \,,
        \end{equation}
    where $D_\chi=n_\tau\cdot\nabla_\chi$ and $\mathcal{H}(Z)$ is the promotion of the Hamiltonian function to an operator acting on $Z$.
    To obtain a specific expression for $\mathcal{H}(Z)$ we note that $\partial_{\xi_c} Z=\kappa Z\partial_{\xi_c}\Omega$, so that one can make the identification: $\rho_c\to(\partial_{\xi_c}Z)/\kappa Z$.
    Indeed, in quantum mechanics, the momentum operator associated to a given physical property is prescribed as the variation with respect to that physical property.
   
    Eq. (\ref{GS}) describes how the distribution $Z$ of mechanical properties $\xi$ evolves over the coordinates $\chi$.
    Even more, since it relates the variation of $Z$ relative to $\xi$ with that relative to $\chi$, it also encodes the dynamics of $\xi$ with respect to $\chi$.
    However, one may be interested in the \emph{proper time} propagation of $Z$.
    In that case, we denote by $\tau$ an affine parameter of the resulting vector field $\nabla_\chi Z$, with local direction $n_\tau$, so that equation (\ref{GS}) becomes:
        \begin{equation}\label{tGS}
         \partial_\tau Z = \kappa\,\mathcal{H}(Z) \,.
        \end{equation}

\section{Synthetic systems}\label{SS}
    In the previous section we built a framework to regard mechanics as an epistemological theory in rather abstract terms and used statistical and quantum mechanics to motivate certain assumptions.
    In what follows we will develop the ideas introduced in the previous sections by studying different specific instances of the propagation (\ref{GS}) to describe a diversity of systems.
    Instead of departing from a fixed action, we shall assume the form of the partition $Z$ and generator $\mathcal{H}$ and investigate its dynamical behavior.
    This requires us to focus on a class of systems that can be put in the form (\ref{canon1}), we refer to them as \emph{synthetic systems}.
     
    We consider that a synthetic system is given by the triple $\{\chi,\xi,Z\}$, such that $\chi$ is the set of spacetime coordinates, $\xi$ its mechanical properties and $Z=\exp\left[ \Theta(\chi,\kappa) + \kappa\,\Omega(\chi,\xi) \right]$ the partition describing how the observations of $\xi$ distributes over spacetime.
    We also assume that they satisfy eq. (\ref{GS}).
   
    The dynamics of a given system is specified thusly by the relation between the variations of $Z$ with respect to $\chi$ and $\xi$.
    Then, through eq. (\ref{GS}), the $\xi$-part and the $\chi$-part of the propagation describing a system have been distinguished.
    To have richer structures, in general, we may allow the space of parameters $\chi$ and the space of properties $\xi$ to be geometric or to realize representation of groups.
    That is a key feature of distinguishing the differential structures in each space.
    Such a distinction may be exploited in modern geometric formulations of physics.
     
    \subsection{Some simple examples}
    To connect this equation with known quantum systems, let us assume that $\kappa=i/\hbar$.
   
    The simple case of a \emph{non-relativistic free particle} corresponds to $\xi=\left\{\phi\right\}$ and $\chi=\left\{t\right\}$, so that equation (\ref{GS}) becomes:
        \begin{equation}\label{FP}
         i\hbar\,\partial_t Z = -\frac{\hbar^2}{2m}\,\partial_{\phi}^2 Z \,,
        \end{equation}
    where $m$ is the mass of the particle.
    Since (\ref{FP}) is a complex equation, the proposed form (\ref{canon1}) leads to two equations:
        \begin{equation}
         \partial_t\Omega + (\partial_\phi\Omega)^2 = 0 \,,
         \quad
         {\rm i}\hbar\left(2\,\partial_t\Theta + \partial^2_\phi\Omega\right) = 0 \,.
        \end{equation}
    The first prescribes that $\Omega$ satisfies the corresponding Hamilton-Jacobi equation, whereas the second relates both functions, $\Theta$ and $\Omega$.
    Upon solving them one obtains:
        \begin{equation}
         \Omega = \frac{m(\phi-\phi_0)^2}{2t} \,,
         \quad\text{and}\quad
         \Theta = -\frac{1}{2}\,\ln\left(t \right) \,.
        \end{equation}
   
    Similarly, the corresponding form of (\ref{GS}) for the \emph{simple harmonic oscillator} is:
        \begin{equation}
         i\hbar\,\partial_t Z = -\frac{\hbar^2}{2m}\,\partial_{\phi}^2 Z + \frac{m\omega^2}{2}\phi^2 Z \,.
        \end{equation}
    This is solved by (\ref{canon1}) with:
        \begin{equation}
         \Omega = \frac{m\omega}{2}\left[(\phi_0^2+\phi^2)\cot(\omega t)-2\phi_0\,\phi\csc(\omega t))\right] \,,
         \end{equation}
        \begin{equation}
         \Theta = -\frac{1}{2}\,{\rm ln}\left[\sin(\omega t) \right] \,.
        \end{equation}
    Again, $\Omega$ satisfies the corresponding Hamilton-Jacobi equation and $\Theta$ relates to $\Omega$ through the equation $2\,\partial_t\Theta + \partial^2_\phi\Omega = 0$.
   
    In the two previous cases we considered that the mechanical property $\phi$ is the position of the corresponding particle.
    This reveals an epistemological assumption behind those systems: that the position is a property of the system that can be dynamically described over a single parameter identified with time.
    Thus, the position is assumed to be distributed in a point-wise manner over the only coordinate.
    One can recognize this as the notion behind the concept of a particle as an ideal in mechanics.
    At the same time, this commences to show the plasticity of the framework presented here.
    Further below, we shall exploit this flexibility.
   
    The previous examples are instances of a generator of the form
        \begin{equation}
         \mathcal{H}=-\frac{a}{\kappa^2}\partial_{\phi}^2 Z + V(\phi)Z \,,
        \end{equation}
    that, for $\kappa=i/\hbar$ and a single $\chi=\left\{t\right\}$, yields a complex propagation (\ref{GS}) that separates into:
        \begin{equation}
         \partial_t\Omega + a(\partial_\phi\Omega)^2 + V(\phi) = 0 \,,
         \quad
         {\rm i}\hbar\left(\partial_t\Theta + a\,\partial^2_\phi\Omega\right) = 0 \,.
        \end{equation}
    Note that it realizes a correspondence principle in the form of the first equation of the Hamilton-Jacobi type and the irrelevance of the second one in the limit $\hbar\to0$.
   
    Before moving on, it is important to recall that the proposed form (\ref{canon1}) yields an effective description resulting from a specific realization of the correspondence principle.
    Furthermore, the choice of a particular form of equation (\ref{GS}) yields the dynamical law of such a description; in this case, the non-relativistic Hamilton-Jacobi formulation of classical mechanics from the Schr\"odinger formulation of quantum mechanics.
    However, there is another simple and interesting case where $\kappa=k\in\mathbb{R}$.
    Now, equation (\ref{GS}) is purely real.
    For example, analogously to the free particle, we have the propagation:
        \begin{equation}
         \partial_t Z = k\,\nabla^2_\xi Z \,,
        \end{equation}
    which corresponds to the \emph{diffusion equation} \cite{m21}.
    Further refinements of $\mathcal{H}$ allow us to produce more complicated equations like the \emph{Fokker-Planck equation} that are useful when dealing with evolving probabilities in statistical mechanics \cite{r84}.
    Note that, as in previous cases, in this case, the mechanical
    properties $\xi$ are commonly the position of the particles.
    However, more diverse physical quantities, such as fields, may be employed.
   
    \subsection{An approach based on a quadratic form}\label{Quad}
    Let us proceed to a larger parameter space $\chi=(x_0,x_1,\dots,x_d)$, while also considering a more pervasive, yet simple case to build upon: a second order $\mathcal{H}$, i.e. quadratic in the impetus.
    Suppose that there is still only one mechanical property $\xi=\{\phi\}$ and the propagation of the corresponding partition is ruled by:
        \begin{equation}\label{Qh}
         \mathcal{H}(Z) = a\,\rho^2(Z) + b\,\rho(Z) + c\,Z \,,
        \end{equation}
    where, in this case, the operator $\rho=\frac{1}{\kappa}\partial_\phi$, and the coefficients $a$, $b$ and $c$ may be functions of $\phi$ and even $\chi$.
   
    The propagating equation (\ref{GS}) becomes:
        \begin{equation}
         \frac{1}{\kappa}D_\chi Z = \frac{a}{\kappa^2}\,\partial_\phi^2Z + \frac{b}{\kappa}\,\partial_\phi Z + c\,Z \,,
        \end{equation}
    that, upon considering a partition of the form (\ref{canon1}), with $\kappa={\rm i}/\hbar$, separates into:
        \begin{equation}\label{QS}
         D_\chi\Omega + a\,(\partial_\phi\Omega)^2 - b\,\partial_\phi\Omega + c = 0 \,,
        \quad
         {\rm i}\hbar\left(D_\chi\Theta + a\,\partial^2_\phi\Omega\right) = 0 \,.
        \end{equation}
   
    Note that, in the limit $\hbar\to0$, the imaginary part (second equation) is null and, with it, the contribution of $\Theta$ to the dynamics.
    On the other hand, the first equation becomes a temporal Hamilton-Jacobi equation for $\Omega$ in terms a particular coordinate, let us say $\chi_0=t$, with an extra term that can be interpreted as a coordinate-dependent potential added to $c$.
    For example, for a (pseudo-) Euclidean spacetime, i.e. $\chi=(t,\pm x)$, one obtains:
        \begin{equation}
         \partial_t\Omega = -a\,(\partial_\phi\Omega)^2 + b\,\partial_\phi\Omega - c \mp D_x\Omega \,,
        \end{equation}
    where $D_x=n_x\cdot\nabla_x$ is the projection of the gradient to the spacial portion of the spacetime.
    This is illustrative of the possibilities of this approach since we have obtained a (seemingly external) contribution to the potential of a system evolving in time only.
    The extra term in the potential is not evident in the original statement of $\mathcal{H}$, although one can readily observe that it derives from the structure of spacetime.

    A specially relevant instance of \ref{Qh} occurs when $a=\frac{1}{2}$, $b=A(\chi)$ and $c=V(\phi)$.
    Then, eqs. (\ref{QS}) takes the form:
        \begin{equation}\label{QS}
         D_\chi\Omega + (\partial_\phi\Omega)^2 - A\,\partial_\phi\Omega + V = 0 \,,
        \quad
         {\rm i}\hbar\left(D_\chi\Theta + \partial^2_\phi\Omega\right) = 0 \,.
        \end{equation}
    In the limit $\hbar\to0$, these yield the dynamics of a field  with generator $\mathcal{H}=\frac{1}{2}\rho^2+A\rho+V$.
    The resulting field may be described otherwise by:
        \begin{equation}
         \nabla^2_\chi\phi = \partial_\phi V - \nabla_\chi A \,.
        \end{equation}
    This is an equation for $\phi(\chi)$, where $V(\phi)$ may be regarded as a potential and $A(\chi)$ as a shift in the momentum.    
   
    \subsection{Further examples}
    Now that we have a general view of the case where $\mathcal{H}$ is of second order for a single property, let us explore some specific instances.
   
    A free \emph{massive scalar field} $\xi=\{\phi\}$ whose dynamics couples to the spacetime coordinates $\chi$ through eq. (\ref{tGS}) is described by:
        \begin{equation}
         \mathcal{H}(Z) = \frac{1}{\kappa^2}\partial_\phi^2Z + m^2\phi^2 Z \,.
        \end{equation}
    Then its propagation separates into the equations:
        \begin{equation}\label{KG}
         D_\chi\Omega + (\partial_\phi\Omega)^2+m^2\phi^2 = 0 \,,
        \quad
         {\rm i}\hbar\left(D_\chi\Theta + \partial^2_\phi\Omega\right) = 0 \,.
        \end{equation}
       
    In the limit $\hbar\to0$, only the generalized Hamilton-Jacobi equation (leftmost eq. in (\ref{KG})) contributes to the dynamics.
    Even more, the first equation in (\ref{KG}) is the generalized Hamilton-Jacobi equation for a Hamiltonian $\mathcal{H}=\rho^2+m^2\phi^2$ that, in turn, yields (macroscopic) dynamical equations for $\phi$ equivalent to $\nabla^2_\chi\phi+m^2\phi=0$.
   
    For a Minkowski parameter space, $\chi=(ct,x)$, one has that $D_\chi=\frac{1}{c}\,\partial_t - D_x$.
    For a single space coordinate, the solution of eqs. (\ref{KG}) is:
        \begin{eqnarray}\nonumber
         \Omega &=& \frac{m}{2}\,\phi^2\,\tan\left[m(x-f_1(ct+x))\right]+\phi\,f_2(ct+x)\,\sec\left[m(x-f_1(ct+x))\right]  \\
            && + \frac{1}{2}\int_1^x f_2(ct+x)^2\,\sec^2\left[m(z-f_1(ct+x))\right]\,dz + f_0(ct+x) \,,
        \end{eqnarray}
        \begin{equation}
         \Theta = f_3(ct+x) - \ln\left(\cos\left[m(x-f_1(ct+x))\right]\right) \,,
        \end{equation}
    where $f_1$, $f_2$, and $f_3$, are differentiable functions of $ct+x$.
   
    Another case we wish to explore is that of a property $\xi=\{\phi\}$ with dynamics ruled by:
        \begin{equation}
         \mathcal{H}(Z) = \frac{m}{\kappa}\,\phi\,\partial_\phi Z \,.
        \end{equation}
    Then eq. (\ref{GS}) separates into:
        \begin{equation}\label{Dr}
         D_\chi\Omega + m\,\phi\,\partial_\phi\Omega = 0\,,
        \quad
         {\rm i}\hbar(D_\chi\Theta) = 0 \,.
        \end{equation}
    Once more, in the limit $\hbar\to0$, only the generalized Hamilton-Jacobi (leftmost eq. in (\ref{Dr})) contributes to the dynamics.
    Even more, the first equation in (\ref{Dr}) is the generalized Hamilton-Jacobi equation for a Hamiltonian $\mathcal{H}=m\,\rho\,\phi$ that, in turn, yields the macroscopic dynamical equation $n_\tau\cdot\nabla_\chi\phi+m\phi=0$.
   
    For a single space coordinate in a Minkowski geometry, a particular solution of eqs. (\ref{Dr}) is:    
        \begin{equation}
         \Omega = f_2\left(ct-\frac{1}{\mu}\ln(\phi),x+\frac{1}{\mu}\ln(\phi)\right) + f_0(ct+x) + c_0\,,
        \end{equation}
        \begin{equation}
         \Theta = f_1(ct+x)
        \end{equation}
    where $f_2$ is an arbitrary differentiable two-variable function, while $f_0$ and $f_1$ are a differentiable functions of $ct+x$, and they hold no relation with those in the previous example.
   
    Finally, consider the case where we have two mechanical properties $\xi=\{\phi,\psi\}$ so that their dynamics is ruled by:
        \begin{equation}\label{MC}
         \mathcal{H}(Z) = \frac{1}{\kappa^2}\,\partial^2_\phi Z + \left(\frac{1}{\kappa}\partial_\psi-\phi\right)^2 Z + V(\phi,\psi)\,Z \,,
        \end{equation}
    This modification in the variation with respect to the mechanical property $\psi$ is commonly known as a minimal coupling between said variable and $\phi$.
    Upon considering a partition of the form (\ref{canon1}), the propagation described by eq. (\ref{GS}), corresponding to (\ref{MC}) separates into:
        \begin{equation}
         D_\chi\Omega + (\partial_\phi\Omega)^2 + (\partial_\psi\Omega+\phi)^2 + V(\phi,\psi) = 0 \,,
        \end{equation}
        \begin{equation}
         {\rm i}\hbar\left(D_\chi\Theta + \partial^2_\phi\Omega + \partial^2_\psi\Omega\right) = 0 \,.
        \end{equation}
    The first is indeed the Hamilton-Jacobi equation for the minimally coupled fields and it is the relevant equation in the limit $\hbar\to0$.
    Notably, if the variation of the principal function with respect to $\phi$ is small so that $\partial_\phi\Omega\approx0$, then one has only the dynamics of $\psi$ minimally coupled to an external field $\phi$:
        \begin{equation}
         D_\chi\Omega + (\partial_\psi\Omega+\phi)^2 + V(\phi,\psi) = 0 \,.
        \end{equation}
    On the other hand, if the variation of the principal function with respect to $\psi$ is small so that $\partial_\psi\Omega\approx0$, then one has the dynamics of a massive field:
        \begin{equation}
         D_\chi\Omega + (\partial_\phi\Omega)^2 + \phi^2 + V(\phi,\psi) = 0 \,,
        \end{equation}
    In all cases $V$ remains a potential term.

\section{Complex action}\label{CA}
    Consider now a complex principal function $\Omega = \Omega_m + {\rm i}\,\Omega_g$, whose partition is given by:
        \begin{equation}\label{ComPart}
         Z = \exp\left[ \Theta + \kappa\,(\Omega_m + {\rm i}\,\Omega_g) \right] \,.
        \end{equation}
    Again, $Z$ satisfies eq. (\ref{GS}), $\kappa={\rm i}/\hbar$ and $\bar{\mathcal{H}}(Z)$ is a second order differential operator with respect to the mechanical properties $\xi$.
    Further, we suppose that the space of $\xi$ is Cartesian, so that $\mathcal{H}(Z)=\frac{a_2}{\kappa^2}\nabla^2_\xi Z+\frac{a_1}{\kappa}\nabla_\xi Z+a_0 Z$, where the coefficients $a_0$, $a_1$ and $a_2$ are generally functions of $\xi$ and even $\chi$.
    Then, separating the propagation into its real and imaginary parts yields:
        \begin{equation}
         D_\chi\Omega_m + a_2(\nabla_\xi\Omega_m)^2 - a_1 \nabla_\xi\Omega_m + a_0 = a_2(\nabla_\xi\Omega_g)^2 - a_2\hbar\nabla^2_\xi\Omega_g \,,
        \end{equation}
        \begin{equation}
         D_\chi\Omega_g + (2a_2\nabla_\xi\Omega_m-a_1)(\nabla_\xi\Omega_g) = \hbar(D_\chi\Theta+a_2\nabla^2_\xi\Omega_m) \,.
        \end{equation}
    These equations describe the evolution of $\Omega_m$ and $\Omega_g$, respectively.
    In the limit $\hbar\to0$, the evolution of $\Omega_m$ is ruled by a Hamilton-Jacobi equation with the extra term $a_2(\nabla_\xi\Omega_g)^2$.
    In such a macroscopic limit, this term can be regarded as a potential external to $\Omega_m$.
   
    Let us build on this case by imposing further conditions to the system.
    Specifically, we wish to distinguish a special class of mechanical properties, $\phi$, that modulate and constrain the rest, $\psi$.
    Therefore, suppose that the whole set of properties $\xi$ is separated as follows: $\xi=\{\phi_1,\dots,\phi_m,\psi_0,\dots,\psi_{n-m}\}$, such that $\Omega_m=\Omega_m(\phi,\psi,\tau)$, whereas $\Omega_g=\Omega_g(\phi,\tau)$, i.e. the real part of the principal function does not depend on $\psi$.
    Further, for simplicity, suppose that $\mathcal{H}=\frac{b_2}{\kappa^2}\nabla^2_\psi Z+\frac{b_0}{\kappa^2}\nabla^2_\phi Z$, where $b_2$ and $b_0$ are functions of $\xi$ and $\chi$.
    Then, eq. (\ref{GS}) separates into:
        \begin{equation}
         D_\chi\Omega_m + b_2(\nabla_\psi\Omega_m)^2 + b_0(\nabla_\phi\Omega_m)^2 = b_0(\nabla_\phi\Omega_g)^2 - b_0 \hbar \nabla^2_\phi\Omega_g \,,
        \end{equation}
        \begin{equation}
         D_\chi\Omega_g + 2b_0(\nabla_\phi\Omega_m)(\nabla_\phi\Omega_g) = \hbar\left(D_\chi\Theta + b_2\nabla^2_\psi\Omega_m + b_0\nabla^2_\phi\Omega_m\right) \,.
        \end{equation}  
   
    Note that, even if one proposes to further separate the principal function in the form of $\Omega_g=\Omega_g(\phi,\tau)$ and $\Omega_m=\Omega_m(\psi,\tau)$, their evolution remains coupled as follows:        
        \begin{equation}\label{GMfree1}
         D_\chi\Omega_m + b_2(\nabla_\psi\Omega_m)^2 - b_0(\nabla_\phi\Omega_g)^2 = -\hbar\,b_0 \nabla^2_\phi\Omega_g \,,
        \end{equation}
        \begin{equation}\label{GMfree2}
         D_\chi\Omega_g = \hbar\left(D_\chi\Theta + b_2\nabla^2_\psi\Omega_m\right) \,.
        \end{equation}
    In the limit $\hbar\to0$, the term $-b_0(\nabla_\phi\Omega_g)^2$ acts as a potential in the dynamics of $\psi$, while $\Omega_g$ is conserved in the direction $n_\tau$.
    Also, special cases can be obtained if, for instance, $\Omega_g=0$, then one has the case of a free dynamics of $\psi$.
    On the other hand, if $\Omega_m=0$, the equations ruling the dynamics of $\phi$ are:
        \begin{equation}
         (\nabla_\phi\Omega_g)^2 = \hbar\nabla^2_\phi\Omega_g \,,
         \qquad
         \partial_\tau\Omega_g = \hbar\,D_\chi\Theta \,;
        \end{equation}
    that, in the macroscopic limit, yield a dynamics of $\phi$ with a null Hamiltonian and a conserved principal function.
   
    In any case, as an illustrative example, instead of the eq. (\ref{GS}), consider eq. (\ref{tGS}).
    That is, instead of discovering the path in the space of $\chi$, let us focus on how the system evolves along the affine parameter $\tau$.
    Solving the system (\ref{GMfree1})-(\ref{GMfree2}), with $D_\chi\to\partial_\tau$, for just one $\psi$ and a single $\phi$ results in:
        \begin{equation}
         \Omega_m(\psi,\tau) = \frac{\psi^2}{b_2(c_3\hbar+2\tau)} + \frac{c_2^2\,\hbar^2}{2}b_0\,\tau + c_5\,,
        \end{equation}
        \begin{equation}
         \Omega_g(\phi,\tau) = f_0(\tau) - \hbar\,{\ln}(\cosh(c_2\phi+c_1)) \,,
        \end{equation}
        \begin{equation}
         \Theta(\tau) = -\frac{1}{2}\ln(2\tau+c_3\hbar) + \frac{f_0(\tau)}{\hbar} + c_4\,,
        \end{equation}
    where $f_0$ is a function of $\tau$, which in turn yields:
        \begin{equation}
         Z(\phi,\psi,\tau) = \frac{\cosh(c_2\phi+c_1)}{\sqrt{c_3\hbar+2\tau}}\,\exp\left(\frac{\rm i}{\hbar}  \left(\frac{\psi^2}{b_2(c_3\hbar+2\tau)} + c_5\right) + \frac{{\rm i}\,\hbar\,c_2^2}{2}b_0\,\tau + c_4\right) \,,
        \end{equation}
    where $c_i$, $i=1,\dots,5$, are constants.
   
    \subsection{An interesting complex case}
    Based on the above results, let us consider now a system with two properties $\xi=\{\phi,\psi\}$, a partition of the form (\ref{ComPart}), with $\Theta=\Theta(\chi)$, $\Omega_g=\Omega_g(\phi,\chi)$ and $\Omega_m=\Omega_m(\psi,\phi,\chi)$; and a dynamics given by:
        \begin{equation}\label{TM}
         \mathcal{H}(Z) = \frac{1}{\kappa}\left(\partial_\phi + \partial_\psi \right)^2Z \,.
        \end{equation}
    Then, equation (\ref{GS}) separates into:        
        \begin{equation}\label{TCm}
         D_\chi\Omega_m + (\partial_\psi\Omega_m + \partial_\phi\Omega_m)^2 = (\partial_\phi\Omega_g)^2 - \hbar\,\partial^2_\phi\Omega_g \,,
        \end{equation}
        \begin{equation}\label{TCg}
         D_\chi\Omega_g + 2\left(\partial_\psi\Omega_m + \partial_\phi\Omega_m\right)\partial_\phi\Omega_g = \hbar\left(D_\chi\Theta + \partial^2_\psi\Omega_m + 2\partial_\psi\partial_\phi\Omega_m + \partial^2_\phi\Omega_m\right) \,.
        \end{equation}
       
    The evolution of $\Omega_m$ is quadratic in its corresponding impetus, $\partial_\psi\Omega_m$, although it is coupled to $\partial_\phi\Omega_m$.
    There is also an extra potential-like term depending on $\Omega_g$.
    On the other hand, the behavior of $\Omega_g$ is somewhat different.
    It is linear in the corresponding impetus $\partial_\phi\Omega_g$ and couples to the impetus relative to $\Omega_m$.
    Further, the evolution of $\Omega_g$ is directly related to the spread of the distribution over the parameters $\chi$, ruled by $\Theta(\chi)$.
 
    \subsubsection{The macroscopic limit}
    Consider now eq. (\ref{TCm}) in the limit $\hbar\to0$.
    For simplicity, let us also focus on the dynamics relative to the the affine parameter $\tau$, that is with $D_\chi\to\partial_\tau$.
    Let us also define $V_g=(\partial_\phi\Omega_g)^2$ and $A=\partial_\phi\Omega_m$, with the additional assumption that it does not depend on $\psi$, that is, $A=A(\phi,\tau)$.
    Note that $V_g=V_g(\phi,\tau)$ from the initial assumption about $\Omega_g$.
    Then one obtains that:
        \begin{equation}\label{cTCm}
         \partial_\tau\Omega_m + (\partial_\psi\Omega_m + A(\phi,\tau))^2 = V_g(\phi,\tau) \,.
        \end{equation}
   
    This describes the dynamics of a field $\psi$, subject to another $\phi$.
    The latter acts on the former by means of a potential term $V_g$ and a shift $A$ in the impetus; both functions of $\phi$ and $\tau$.
    Otherwise, $\psi$ would be a free field.
    The corresponding principal function is given by:
        \begin{equation}
         \Omega_m(\psi,\phi,\tau) = C_0\,\psi + f_1(\phi) + \int V_g(\phi,\tau)-(A(\phi,\tau)+C_0)^2\,{\rm d}\tau \,,
        \end{equation}
    where $C_0$ is a constant, since $\partial_\phi\Omega_m$ does not depend on $\psi$, and $f_1$ is a function of $\phi$ only.
    Also, one can see that $\partial_\psi\Omega_m=C_0$, whereas $A$ satisfies:
        \begin{equation}\label{Gis1}
         \partial_\tau A + 2\,(A+C_0)\partial_\phi A = \partial_\phi V_g \,.
        \end{equation}
    Thus, the contribution $A=\partial_\phi\Omega_m$ to the impetus of $\psi$ seems to have its own  evolution analogously to a principal function, not evident in (\ref{TM}).
   
    Now, let us turn to eq. (\ref{TCg}).  
    In the limit $\hbar\to0$, it can be written as:
        \begin{equation}\label{Gis2}
         \partial_\tau\Omega_g + 2\,(A+C_0)\partial_\phi\Omega_g = 0 \,.
        \end{equation}
    Note that the system (\ref{Gis1})-(\ref{Gis2}) depends only on $\phi$ and $\tau$.
    Actually, eq. (\ref{Gis2}) implies the condition:
        \begin{equation}\label{Fhi}
         \partial_\tau\phi + 2(\partial_\phi\Omega_m + \partial_\psi\Omega_m) = 0 \,,
        \end{equation}
    which means that the variation of $\phi$ along $\tau$ is proportional to the (shifted) impetus of $\psi$.
    It also implies that $\partial_\phi A=0$, and:
        \begin{equation}
         \Omega_g = f_2\left(\tau - \frac{\phi}{2(A+C_0)}\right) \,,
        \end{equation}
    where $f_2$ is a single variable function; whereas eq. (\ref{Gis1}) becomes $\partial_\tau A = \partial_\phi V_g$, and:
        \begin{equation}\label{geq}
         \frac{1}{2}\partial^2_\tau\phi + \partial_\phi V_g = 0 \,,
        \end{equation}
    which is already a dynamical equation for $\phi$, with potengial $V_g$.
   
    As a brief summary of the features of this system consider that, through a partition (\ref{ComPart}), with the provision that $\partial_\psi\Omega_g=0$, whose propagation is ruled by eq. (\ref{tGS}), with $\mathcal{H}$ given by (\ref{TM}), we have obtained a model for the dynamics of two physical characteristics, $\psi$ and $\phi$.
    While they have different types of evolution, they are distinctively coupled, even in the limit $\hbar\to0$.
    In this macroscopic limit, on the one hand, the dynamics of $\psi$ is that of a field minimally coupled to $A=\partial_\tau\phi$ and evolves under the influence of a potential $V_g(\phi,\tau)$;
    while, on the other hand, the dynamics of $\phi$ depends on both, the momentum flux of $\psi$ and $\phi$ itself, rahter than just the squared of the latter.
   
    Even more, other cases can be obtained by imposing further conditions on the system.
    For example:
        \begin{itemize}
         \item If $\partial_\phi\Omega_m=0$, there is no coupling with the impetus of $\psi$.
         \item If $\partial_\phi\Omega_g=0$, there is no external potential $V_g$ affecting the dynamics of $\psi$. This and the previous case together lead to a free field dynamics of $\psi$.
         \item If $\Omega_m=0$, then $\Omega_g$ is not necessarily null, but it must be a conserved quantity along $\tau$ that depends only on parameters $\chi$.
        \end{itemize}
     
    The case $\Omega_m=0$ is far from trivial, since one can still use the thermodynamic relation (\ref{1id}), incorporating the properties of the space of parameters $\chi$.
    Here, with $\Omega_m=0$, the free action $F$ and entropy $\sigma$ become directly tied to the imaginary part $\Omega_g$ of the principal function and the noise term $\Theta$.
    This is analogous to studying the thermodynamics of the respective system.
    For the case at hand, eq. (\ref{1id}) yields:
        \begin{equation}
         F = \Omega_m + {\rm i}\Omega_g + \frac{1}{\kappa}\,\Theta \,.
        \end{equation}
   
    Note that, if $\kappa={\rm i}/\hbar$, the free action $F$ becomes complex in general: $F=\Omega_m+{\rm i}\,(\Omega_g-\hbar\,\Theta)$.
    This complex free action encodes both the dynamical (real) and thermodynamical (imaginary) aspects of the system, consistent with the complex action formalism developed in this section.
   
    If we define the \emph{internal action} as $U:=F-\Omega$ (here we have $\Omega=\Omega_m+{\rm i}\,\Omega_g$), then passing to the thermodynamic potential $U$ description yields the following  equation of state:
        \begin{equation}\label{EoS}
         {\rm d}U = \frac{1}{\kappa}\,{\rm d}\Theta \,.
        \end{equation}
 
    For example, in \cite{j95}, the Einstein's field equations are derived from an equation similar to (\ref{EoS}), where the following identifications is made:
        \begin{equation}
         {\rm d}U = \int \tau T_{ij}\,{\rm d}h^{ij}\,{\rm d}\tau \,,
         \qquad
         {\rm d}\Theta = -\int \tau R_{ij}\,{\rm d}h^{ij}\,{\rm d}\tau \,.
        \end{equation}
    whereas the term $\left(-\frac{1}{2}R+\Lambda\right)g_{ij}$ was obtained from a freedom to add a divergence-less function along with the contracted Bianchi identity.
    However, in the light of the results of this section, if $\Omega_m=0$, then eq. (\ref{EoS}) still holds, where $U=F-{\rm i}\,\Omega_g$, which depends on $\chi$ and $g_{ij}$.
    This allows us to conjecture that they might contribute to such a term.
   
    The fact that the principal function $\Omega=\Omega_m+{\rm i}\,\Omega_g$ is complex gives way to a dynamics of $\phi$ and $\psi$ coupled in a specific way, where the former modifies the evolution of the latter.
    At the same time $\Omega_m$ is distinctly involved in the evolution of $\psi$ and $\Omega_g$ in the evolution of $\phi$.
    Even more, $\Omega_m$ being the real part of $\Omega$, yields $\psi$ as a regular (matter and energy) field;
    whereas, $\Omega_g$ being the imaginary part of $\Omega$, renders $\phi$ as a special type of (thermodynamical) field that connects directly with the generalized entropy $\Theta$ and whose macroscopic behavior is obtained non-trivially by means of thermodynamics, rather than the usual Hamiltonian way.
   
    \subsection{Scalar field in a Robertson-Walker metric}
    An final illustrative application of the ideas presented here is the case of a real scalar field $\psi$ in an isotropic-homogeneous spacetime describable through the scale factor $\phi$.
    Their dynamics with respect to a single time coordinate $t$ is usually derived from a Lagrangian:
    \begin{equation}
     \mathcal{L}= -\frac{1}{8\,\pi G}\left(3\,\phi\,(\partial_t\phi)^2-3\,k\,\phi+\Lambda\phi^3)\right) + \phi^3\left(\frac{1}{2}(\partial_t\psi)^2-V(\psi)\right) \,,
    \end{equation}
    where $G$ is the gravitational constant and $\Lambda$ is the cosmological constant.
    Then we have $\xi=\{\phi,\psi\}$ and $\chi=\{t\}$.
   
    If one wishes to apply (\ref{GS}), a natural choice for $\mathcal{H}$ is the associated Hamiltonian, promoted to an operator acting on partition $Z$:
    \begin{equation}\label{RWr}
     \mathcal{H}(Z) = -\frac{2\,\pi G}{3\,\phi}\,\hbar^2\,\partial^2_\phi Z + \frac{1}{8\,\pi G}(3\,k\,\phi-\Lambda\phi^3)\,Z + \frac{\hbar^2}{2\,\phi^3}(\partial_t\psi)^2 + \phi^3\,V(\psi)\,Z \,,
    \end{equation}
    where we implemented $\rho_\phi\to\partial_\phi Z$ and $\rho_\psi\to\partial_\psi Z$.
    Then, we proceed to separate eq. (\ref{GS}) under the assumption (\ref{canon1}) to yield the system of equations:
        \begin{equation}\label{syst11}
         \partial_t\Omega - \frac{2\,\pi\,G}{3\,\phi}\,(\partial_\phi\Omega)^2 - \frac{3\,k}{8\,\pi\,G}\phi + \Lambda\,\phi^3 + \frac{1}{2\,\phi^3}\,(\partial_\psi\Omega)^2 + \phi^3\,V(\psi)= 0 \,,
        \end{equation}
        \begin{equation}\label{syst12}
         6\,\phi^2\,\partial_t\Theta - 4\,\pi\,G\,\phi\,\partial^2_\phi\Omega + \frac{3}{\phi}\,\partial^2_\psi\Omega = 0 \,.
        \end{equation}
   
    We can observe that eq. (\ref{syst11}) is indeed the dynamical equation of the configuration in the classical regime, once we impose that $\Omega$ be independent of time since, as is well known, the Hamiltonian of the system is null \cite{adm62}.
    This observation, together with eq. (\ref{syst12}), that yields $\Theta=\theta\,t$, for some constant $\theta$.
    The resulting system is:
        \begin{equation}\label{syst13}
         \frac{2\,\pi\,G}{3\,\phi}\,(\partial_\phi\Omega)^2 + \frac{3\,k}{8\,\pi\,G}\phi = \Lambda\,\phi^3 + \frac{1}{2\,\phi^3}\,(\partial_\psi\Omega)^2 + \phi^3\,V(\psi)  \,,
        \end{equation}
        \begin{equation}\label{syst14}
         6\,\theta\,\phi^2 - 4\,\pi\,G\,\phi\,\partial^2_\phi\Omega + \frac{3}{\phi}\,\partial^2_\psi\Omega = 0 \,.
        \end{equation}
    Equation (\ref{syst13}) is the condition that the Hamiltonian is null, while equation (\ref{syst14}) governs the quantum aspects of the distribution $Z$.
   
    Despite the approach described above, we have the alternative of a complex action, as in eq. (\ref{ComPart}), where now we choose $\Omega_g=\Omega_g(\phi,t)$ and $\Omega_m=\Omega_m(\psi,\phi,t)=\Phi(\phi,t)+\Psi(\psi,t)$.
    Let us propose a dynamical operator of the form:
    \begin{equation}
     \mathcal{H}(Z) = -\hbar^2\,\partial^2_\phi Z - \frac{\hbar^2}{8\,\pi\,G\,\phi^3}\,\partial^2_\psi Z + \left(8\,\pi\,G\,\phi^3\,V(\psi) + \frac{\Lambda}{3}\,\phi^2 - k\right)\,Z \,.
    \end{equation}
    Again, eq. (\ref{GS}) under the assumption (\ref{canon1}) separates into:
    \begin{equation}\label{syst21}\footnotesize
     \partial_t\Psi + \frac{1}{16\,G\,\pi\,\phi^3}(\partial_\psi\Psi)^2 + 8\,\pi\,G\,\phi^3\,V(\psi)       = - \partial_t\Phi - (\partial_\phi\Phi)^2 + \frac{\Lambda}{3}\,\phi^2 -k + (\partial_\phi\Omega_g)^2 + \hbar\,\partial^2_\phi\Omega_g \,,
    \end{equation}
    \begin{equation}\label{syst22}
     \partial_t\Omega_g + 16\,(\partial_\phi\Phi)\,(\partial_\phi\Omega_g) = -\hbar\,\left(\partial_t\Theta + \partial^2_\phi\Phi + \frac{1}{16\,\pi\,G\,\phi^3}\,\partial^2_\psi\Psi \right) \,.
    \end{equation}  
   
    Now, we consider the classical limit: $\hbar\to0$ to confirm that we obtain the appropriate dynamical system.
    First, we have the following system of equations:
    \begin{equation}\label{syst23}\small
     \partial_t\Psi + \frac{1}{16\,G\,\pi\,\phi^3}\,(\partial_\psi\Psi)^2 + 8\,\pi\,G\,\phi^3\,V(\psi)
     = - \partial_t\Phi - (\partial_\phi\Phi)^2 - \frac{\Lambda}{3}\,\phi^2 + k + (\partial_\phi\Omega_g)^2 \,,
    \end{equation}
    \begin{equation}\label{syst24}
     \partial_t\Omega_g + 16\,(\partial_\phi\Phi)\,(\partial_\phi\Omega_g) = 0 \,.
    \end{equation}
    Next, note that, although the right-hand side of eq. (\ref{syst23}) is non-null, it does not depend on $\psi$.
    Then, the corresponding dynamical equation for the scalar field $\psi$ in the classical regime is indeed:
    \begin{equation}
     \partial^2_t\psi + \frac{3}{\phi}\,(\partial_t\phi)\,(\partial_t\psi) + \partial_\phi V = 0 \,.
    \end{equation}
    
    On the other hand, similar to the complex action treatment before, we can solve eq. (\ref{syst23}) for $\partial_\phi\Phi$ and substitute it in eq. (\ref{syst24}).
    However, if we also impose the condition:
    \begin{equation}
     (\partial^2_\phi\Omega_g)^2 = \frac{\Omega_g}{4\,\phi} + (\partial_t\Phi + \partial_t\Psi) \,,
    \end{equation}
    then one obtains that:
    \begin{equation}
     \partial_t\Omega_g + \sqrt{\frac{\Omega_g}{\phi} + \frac{\Lambda}{3}\,\phi^2 - k -4\left(\frac{1}{16\,G\,\pi\,\phi^3}\,(\partial_\psi\Psi)^2 + 8\,\pi\,G\,\phi^3\,V(\psi)\right)}\,\partial_\phi\Omega_g = 0 \,. 
    \end{equation}
    This in turn yields the following dynamical equation in the classical regime:
    \begin{equation}
     2\,\phi\,\partial^2_t\phi + (\partial_t\phi)^2 + k = \Lambda\,\phi^2 + \left(\frac{1}{16\,G\,\pi\,\phi^3}\,(\partial_\psi\Psi)^2 + 8\,\pi\,G\,\phi^3\,V(\psi)\right) \,,
    \end{equation}
    which is the dynamical equation for the scale factor $\phi$ in the classical regime.
   
    Before concluding, let us note that the assumption of the form (\ref{canon1}) is a highly restrictive imposition.
    Further work shall be oriented to enhance the results obtained here for the general case:
        \begin{equation}\label{gral}
         Z = \exp\left( \Theta(\xi_c,\chi,\kappa) + \kappa\,\Omega(\xi_c,\chi) \right) \,.
        \end{equation}
 
\section{Conclusions}\label{CC}
    This work investigates an epistemological interpretation of mechanics.
    Motivated by the formal coincidences between quantum mechanics and statistical mechanics, in terms of the path integral and partition function analogy, we propose that the dynamical problem of a system consists in finding a distribution of its mechanical properties over a space of parameters with a given local structure.
    Notably, in the case of most interest to physics those parameters are the coordinates of a spacetime with a (pseudo-) Riemannian local structure.
    Simpler cases include the Lorentzian case, used in quantum field theory, and the Euclidean case, used in statistical mechanics.
   
    Furthermore, the epistemological generalization of mechanics implies that the observed values of mechanical properties act as constrictions on their distributions.
    Thus, the dynamical problem must satisfy observations as boundary values.
    The distribution is readily connected to the partition function as a functional integral in (\ref{parti}) and a probabilistic interpretation is given in the standard fashion.
   
    The first relevant result is eq. (\ref{1id}) that is analogous to the fundamental thermodynamic relation in statistical mechanics.
    Then, the correspondence principle is used to propose the specific form of the partition shown in eq. (\ref{canon1}).
    The dynamical problem is then fixed as a generalization of the Shcr\"odinger equation as depicted in eq. (\ref{GS}).
    This equation describes the operator $\mathcal{H}$ as the generator of translations of the partition over a spacetime.
   
    After that initial exploration, in Section (\ref{GME}), a concrete proposal was presented for a certain type of systems that depict the epistemological features in a simple way.
    We dubbed those as synthetic systems, composed of a set of mechanical properties, a set of spacetime coordinates and a partition function describing the distribution of the former over the latter.
    Additionally, we demanded that the partition be of the form (\ref{canon1}), satisfying eq. (\ref{GS}).
    Departing from these assumptions, we progressed in an increasing complexity of instances of synthetic systems.
    The idea being that eq. (\ref{GS}) relates differential structures on mechanical properties (internal) and spacetime (external).
    Here, the external structure is $D_\chi Z$, the gradient of the partition in the direction of greatest increase defined by the very same gradient;
    whereas the internal structure is given by $\mathcal{H}(Z)$, a differential operator acting on the partition, that can be interpreted as the generator of spacetime translations.

    This distinction between both sets of variables is greatly useful, since it allows to deal separately with their additional structures such as geometric spaces.
    Specifically, one can be used to prescribe the base spacetime of interest, whereas the other can be used to prescribe the relations among the fields that evolve on said spacetime.
    In algebraic treatments, this distinction yields spacetime and internal symmetries, respectively.
   
    Next, in Section \ref{SS}, the formal proposal is connected to several systems of interest in physics.
    In particular, for a single spacetime parameter, i.e. a single temporal coordinate, and a second order $\mathcal{H}(Z)$, one obtains the usual non-relativistic quantum dynamical law in the Schr\"odinger formulation.
    This corresponds to considering the spatial position as the mechanical property to be distributed over the time coordinate.
    Upon considering a greater number of spacetime coordinates, one may use this framework to study field theories.
    In this case, the field is readily identified with the mechanical property distributed over spacetime.
    It is worth noticing that the physical properties considered here form a Cartesian space.
    However, other cases may be investigated; most notably, the case where the physical properties yield a (non-commutative) group representation.
   
    Finally, in Section \ref{CA}, the initial assumption (\ref{canon1}) is generalized so that the principal function is a complex quantity, where only a class of mechanical properties contributes to the extra imaginary part.
    We dubbed these thermodynamical fields since their Hamiltonian dynamics is trivial and yields only the condition of having a conserved principal function.
    However, the fundamental thermodynamic relation (\ref{1id}) can still be used to study their dynamical behavior.
    This points in the direction of other works that regard gravitation as a thermodynamic phenomenon.
    Here, we show it is possible to relate it to an imaginary contribution to the principal function, so that its dynamics is readily connected to the entropy-like quantity $\Theta$.
    The case of a scalar field in a Robertson-Walker metric is formally explored.
   
    We would like to bring to the attention of the reader that, disregarding the epistemological reinterpretation of mechanics proposed here, there is also a particular value of the results presented here in the solution of dynamical problems in statistical and quantum mechanics.
    The idea being that if a constraining observation is known (or prepared), for example as an initial distribution, one may propose a partition of the form (\ref{canon1}) to propagate over spacetime.
    However, in line with results here, one has introduced a noise-controlling term $\Theta$ that counters the tendency of the system to follow an evolution prescribed by $\Omega$.
    To control the relation between these terms, one uses $\kappa$, as well as its limiting values.
    Then, one has effectively introduced two variables: $\Theta$ and $\kappa$.
   
    Future work along the line of this article includes generalizations as well as applications to relevant systems.
    Among the possible generalizations, we highlight that definition (\ref{FrAct}) and assumption (\ref{canon1}) are based on a simple epistemological interpretation of the canonical ensemble in statistical mechanics, but similar interpretations of other ensembles may be explored.
    In particular, a straightforward line goes along the of a general partition of the form (\ref{gral}).
   
    On the other hand, eq. (\ref{GS}) was obtained from the requirement  that it would correspond to a given macroscopic dynamics in the limit $\kappa\to0$.
    However, other equations may be investigated, in particular, those of higher order.
    Also, instead of focusing on features of the principal function $\Omega$, one could impose conditions on the entropy-like function $\Theta$, for example, using any of the entropy functions available in the literature.
   
    As additional applications one may consider spaces of properties $\xi$ with structures more complicated than the Cartesian case, for example, any of the many particle fields carrying group representations.
    The form of the generator $\mathcal{H}(Z)$ can be generalized accordingly.
    Finally, one may further investigate the mechanism of emergent gravity referred here, particularly for concrete instances of synthetic systems.

\end{document}